# Acoustic Nonreciprocity in Coriolis Mean Flow Systems
Masoud Naghdi, Farhad Farzbod
Department of Mechanical Engineering, University of Mississippi, University, MS 38677

**ABSTRACT:** One way to break acoustic reciprocity is to have a moving wave propagation medium. If the acoustic wave vector and the moving fluid velocity are collinear we can use the wave vector shift caused by the fluid flow to break reciprocity. In this paper we investigated an alternative approach in which the fluid velocity enters the differential equation of the system as a cross product term with the wave vector. A circular field where the fluid velocity increases radially has a Coriolis acceleration term. In such a system, the acoustic wave enters from the central wall and exits from the perimeter wall. In this paper, we solved the differential equation numerically and investigated the effect of fluid velocity on the nonreciprocity factor.

## I. INTRODUCTION

There are two technical terms being used sometimes interchangeably for similar yet different phenomena in acoustics; reciprocity and time reversal symmetry. It is most helpful to first make a distinction between these two different terms. Reciprocity states that if there is a source at point A, and we measure pressure at point B, the pressure would be the same as if we have the source at point B and the measurement is performed at point A. Time reversal symmetry, on the other hand, states that if $u(t,x)$ is the solution to the wave equation so is $u(-t,x)$ with $t$ and $x$ being the time and space variable.

Reciprocity in acoustics was first discussed by Helmholtz [1] in his acoustic analysis of open ended pipes. later J.W. Strutt (Lord Rayleigh) [2] generalized this theorem for any linear system including the ones with energy dissipation. More formal proof of this theory in a general case of anisotropic media was presented by Lyamshev [3]. To have reciprocity in an acoustic medium, it is not necessary to preserve energy, i.e. a linear system can have damping in any form while satisfying reciprocity condition [4, 5]. However, the medium should be at rest for the theorem to hold true [6, 7]. One way to break reciprocity is to have a moving medium in the system. Fluery et. al.[8] recently used this effect to devise an acoustic circulator and a conceptual resonator lattice structure [9]. Another approach is feasible when there is an external fields such as a magnetic field. If this interaction manifests itself as a cross product term in the governing differential equation of the system, it can be shown that reciprocity is broken. In the case of electromagnetic waves, magnetic fields fit this criteria. Faraday rotators, with several applications in optical devices and systems, use the interaction of a magnetic field with the propagation medium to achieve non-reciprocity in optics[10]. Kittel[11] showed the possibility the possibility of a similar phenomenon in a ferromagnetic crystal and it was later verified experimentally[12, 13] for surface acoustic waves in aluminum. The drawback of the latter approach, however, is its limited application due to its typical occurrence at high frequencies (10 MHz) and at low levels of nonreciprocity.

Time reversal symmetry can have several meanings [14], in one sense it means that if $u(t,x)$ $t: t_1 \rightarrow t_2$ is the solution to the governing differential equation, then $u(t' = -t, x)$ $t':-t_2 \rightarrow -t_1$ is also a solution. Time reversal symmetry in this sense, among others, depends on the order of time derivatives and the coefficients. For example viscous damping violates the symmetry in time (Fig. 1). It should be noted that time reversal symmetry, here, is different with reversing the time, i.e. going back in time [15]. For example, in a system with viscous damping, if we march back in time, the energy is going back to the system and entropy decreases. This concept of marching back in time, or active time reversal, can be utilized to focus acoustic energy in a very small area [16, 17]. In this sense, any deterministic system has time reversal symmetry.

However, the interest in breaking time reversal symmetry in which *u(t,x)* is the solution to the system unlike *u(-t,x)*, stems from the fact that in these cases, we can have asymmetrical wave propagation. Similar to reciprocity breaking, one way to break time reverse symmetry is through application of an external field. If the external field manifest itself in the differential equation through cross product with the first degree time derivative of a variable (such as displacement or pressure), then we can break time reversal symmetry without detriment to the total energy of the system.

External magnetic field historically played a vital role in both nonreciprocal systems and systems without time reversal symmetry. Interaction between magnetic field and beam of quantum particles (such as electrons) posed by Aharonov and Bohm [18] initiated a lot of research due to its paradoxical effect; a magnetic field isolated inside an infinite cylinder with direction parallel to the axis, can scatter quantum particles striking outside of the cylinder without directly "touching" the magnetic field. This phenomena happens because the magnetic vector potential in Schrödinger's equation depends on the magnetic flux, which itself depends on magnetic field passing through the cylinder due to the Stokes relation. Berry et al [19] studied Aharonov-Bohm effect theoretically and experimentally. In order to observe experimentally the interaction between unobservable topology and wavefront, they use resembling phenomena of surface wave interaction with irrotational vortex. The resemblance between these two phenomena stops at both being unobservable topology. In the surface wave case, the wave front and the irrotational vortex are directly "accessing" each other by one passing through the other. This type of interaction between surface wave and vortices were further examined by Coste et. al. for surface waves in shallow water [20] and by Bernal et. al [21] for vorticity field with zero net circulation. There has been other researches performed in the area of vortex interaction with the acoustic wave with applications in jet noise analysis and solid-fluid interactions. These include other types of vortices such as turbulent vortex [22] vortex ring [23], Rankine vortex [24] and vortex dipole [25].

Breaking reciprocity and/or time reversal symmetry can be achieved in linear systems by the methods discussed before. In a nonlinear system, we have more ways to achieve the same goal of making an acoustic and/or thermal diode. This can be done by using nonlinear granular media [26], combining sonic crystal and nonlinear medium [27], nonlinear lattices coupled together by a harmonic spring [28], by means of bifurcations [29] and combining a linear and nonlinear system [30]. Apart from these techniques, there are some methods to make a device with interesting acoustic properties. For instance, Zhu et al. [31] designed a tunnel with metasurfaces in which a plane wave travels through and exit as a non-plane wave while the plane wave doesn't go through as much if it enters the tunnel from the other side. As Maznev et al. [32] explained, these are not in fact 'diodes' nor they break reciprocity, however, they most certainly have interesting applications.

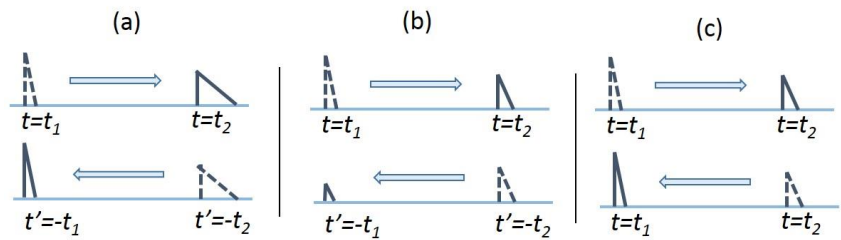

**Figure 1 (a) In a system with time reversal symmetry u(t,x) and u(t'=-t,x) are both solutions. Note that the direction of time is positive in both cases. (b) Wave propagation in a system without time reversal symmetry has asymmetrical wave propagation when we go forward in time; $-t_2$ to $-t_1$ and $t_1$ to $t_2$. (c) Wave propagation in a system without time reversal symmetry, in this case we go forward and backward in time.**

Our approach here to break reciprocity in a linear acoustic system, is to have a moving fluid medium in which the wavevector undergoes Coriolis acceleration [33]. In this way, the medium velocity appears in the wave equation as a cross product term and reciprocity is broken. In the following section, we derive governing equation for such systems and in section 3 we discuss reciprocity. In the final section 4, we present numerical simulation for an example system and wave propagation

## II. WAVE EQUATION IN CORIOLIS MEAN FLOW SYSTEMS

In this section we first derive and then discuss wave equation in a circular field in which the medium velocity increases with the distance from the center of rotation. An example of such system is depicted in Figure 2. This device operates with a wave entering from a wall in the center, and exiting from a wall on the perimeter. Equations of conservation of mass and momentum for such a field, after linearization, can be simplified to:

$$\left(\frac{\partial}{\partial t} + V \cdot \nabla\right) P + \rho C^2 \nabla \cdot (u + V) = 0, \tag{1}$$

$$\rho \frac{\partial}{\partial t}(u + V) + \rho(V \cdot \nabla)u + \rho(u \cdot \nabla)V + \rho(V \cdot \nabla)V + \nabla P = 0. \tag{2}$$

in which $u$ is the acoustic velocity field, $P$ is the acoustic pressure and $V$ is the medium velocity. In the case of circular flow with constant angular velocity, the velocity field is stated as $\boldsymbol{V} = r\boldsymbol{\omega}\hat{\boldsymbol{\theta}}$. It can then be shown that Eqs. 1 and 2, in matrix form, can be simplified as:

$$\frac{\partial}{\partial t}\begin{bmatrix}u_1\\u_2\\P\end{bmatrix} + \begin{bmatrix}-\omega y & 0 & 1/\rho\\0 & -\omega y & 0\\\rho C^2 & 0 & -\omega y\end{bmatrix}\frac{\partial}{\partial x}\begin{bmatrix}u_1\\u_2\\P\end{bmatrix} + \begin{bmatrix}\omega x & 0 & 0\\0 & \omega x & 1/\rho\\0 & \rho C^2 & \omega x\end{bmatrix}\frac{\partial}{\partial y}\begin{bmatrix}u_1\\u_2\\P\end{bmatrix} + \begin{bmatrix}0 & -\omega & 0\\\omega & 0 & 0\\0 & 0 & 0\end{bmatrix}\begin{bmatrix}u_1\\u_2\\P\end{bmatrix} + \begin{bmatrix}-\omega^2 x\\-\omega^2 y\\0\end{bmatrix} = 0 \tag{3}$$

in which $u_1$ and $u_2$ are the acoustic velocity field in $x$ and $y$ directions. The fourth term in Eq. (3) is a cross product between velocity components and the fluid flow rotation. This Coriolis term is responsible for breaking reciprocity.

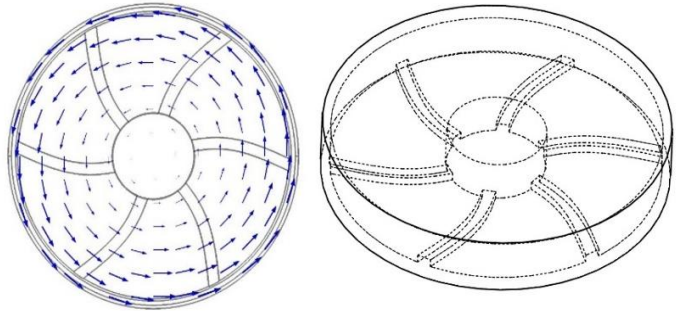

**Figure 2 An example velocity field for a circular flow is depicted on the left. This field can be induced by an example rotating apparatus depicted on the right.**

In order to simplify boundary condition and avoid dealing with both acoustic velocity and pressure at the boundaries, we reduce the number of equation into two equations in terms of $u_1$ and $u_2$. We start by rewriting eq. (3) as:

$$\frac{\partial P}{\partial t} - \omega y \frac{\partial P}{\partial x} + \omega x \frac{\partial P}{\partial y} + \rho C^2 \left(\frac{\partial u_1}{\partial x} + \frac{\partial u_2}{\partial y}\right) = 0, \tag{4}$$

$$\frac{\partial u_1}{\partial t} + \left(-\omega y \frac{\partial u_1}{\partial x} + \omega x \frac{\partial u_1}{\partial y}\right) - \omega u_2 - \omega^2 x + 1/\rho \left(\frac{\partial P}{\partial x}\right) = 0, \tag{5}$$

$$\frac{\partial u_2}{\partial t} + \left(-\omega y \frac{\partial u_2}{\partial x} + \omega x \frac{\partial u_2}{\partial y}\right) + \omega u_1 - \omega^2 y + 1/\rho \left(\frac{\partial P}{\partial y}\right) = 0. \tag{6}$$

Taking partial derivative of equations 4-6 with respect to $x$, $y$ and $t$ give us:

$$\frac{\partial}{\partial x} eq.(4) = \frac{\partial^2 P}{\partial x \partial t} - \omega y \frac{\partial^2 P}{\partial x^2} + \omega \frac{\partial P}{\partial y} + \omega x \frac{\partial^2 P}{\partial x \partial y} + \rho C^2 \left(\frac{\partial^2 u_1}{\partial x^2} + \frac{\partial^2 u_2}{\partial x \partial y}\right) = 0, \tag{7}$$

$$\frac{\partial}{\partial y} eq.(4) = \frac{\partial^2 P}{\partial y \partial t} - \omega \frac{\partial P}{\partial x} - \omega y \frac{\partial^2 P}{\partial x \partial y} + \omega x \frac{\partial^2 P}{\partial y^2} + \rho C^2 \left(\frac{\partial^2 u_1}{\partial x \partial y} + \frac{\partial^2 u_2}{\partial y^2}\right) = 0, \tag{8}$$

$$\frac{\partial}{\partial x} eq.(5) = \frac{\partial^2 u_1}{\partial x \partial t} + \left(-\omega y \frac{\partial^2 u_1}{\partial x^2} + \omega \frac{\partial u_1}{\partial y} + \omega x \frac{\partial^2 u_1}{\partial x \partial y}\right) - \omega \frac{\partial u_2}{\partial x} - \omega^2 + \frac{1}{\rho}\left(\frac{\partial^2 P}{\partial x^2}\right) = 0, \tag{9}$$

$$\frac{\partial}{\partial y} eq.(5) = \frac{\partial^2 u_1}{\partial y \partial t} + \left(-\omega \frac{\partial u_1}{\partial x} - \omega y \frac{\partial^2 u_1}{\partial x \partial y} + \omega x \frac{\partial^2 u_1}{\partial y^2}\right) - \omega \frac{\partial u_2}{\partial y} + \frac{1}{\rho}\left(\frac{\partial^2 P}{\partial x \partial y}\right) = 0, \tag{10}$$

$$\frac{\partial}{\partial t} eq.(5) = \frac{\partial^2 u_1}{\partial t^2} + \left(-\omega y \frac{\partial^2 u_1}{\partial x \partial t} + \omega x \frac{\partial^2 u_1}{\partial y \partial t}\right) - \omega \frac{\partial u_2}{\partial t} + \frac{1}{\rho}\left(\frac{\partial^2 P}{\partial x \partial t}\right) = 0, \tag{11}$$

$$\frac{\partial}{\partial x} eq.(6) = \frac{\partial^2 u_2}{\partial x \partial t} + \left(-\omega y \frac{\partial^2 u_2}{\partial x^2} + \omega \frac{\partial u_2}{\partial y} + \omega x \frac{\partial^2 u_2}{\partial x \partial y}\right) + \omega \frac{\partial u_1}{\partial x} + \frac{1}{\rho}\left(\frac{\partial^2 P}{\partial x \partial y}\right) = 0, \tag{12}$$

$$\frac{\partial}{\partial y} eq.(6) = \frac{\partial^2 u_2}{\partial y \partial t} + \left(-\omega \frac{\partial u_2}{\partial x} - \omega y \frac{\partial^2 u_2}{\partial x \partial y} + \omega x \frac{\partial^2 u_2}{\partial x \partial y}\right) + \omega \frac{\partial u_1}{\partial y} - \omega^2 + \frac{1}{\rho}\left(\frac{\partial^2 P}{\partial y^2}\right) = 0, \tag{13}$$

$$\frac{\partial}{\partial t} eq.(6) = \frac{\partial^2 u_2}{\partial t^2} + \left(-\omega y \frac{\partial^2 u_2}{\partial x \partial t} + \omega x \frac{\partial^2 u_2}{\partial y \partial t}\right) + \omega \frac{\partial u_1}{\partial t} + \frac{1}{\rho}\left(\frac{\partial^2 P}{\partial y \partial t}\right) = 0. \tag{14}$$

Now, in equations 7 and 8, all the terms with $P$ can be replaced with terms containing $u_1$ and $u_2$ from equations 9-14. Consequently we have two equations in terms of $u_1$, $u_2$ and their time and spatial derivatives:

$$-2\omega x \left(\frac{\partial^2 u_1}{\partial y \partial t}\right) + 2\omega y \left(\frac{\partial^2 u_1}{\partial x \partial t}\right) - \omega^2 y^2 \left(\frac{\partial^2 u_1}{\partial x^2}\right) + \omega^2 y \left(\frac{\partial u_1}{\partial y}\right) + 2\omega^2 xy \frac{\partial^2 u_1}{\partial x \partial y} + \omega^2 x \left(\frac{\partial u_1}{\partial x}\right)$$
$$- \omega^2 x^2 \left(\frac{\partial^2 u_1}{\partial y^2}\right) + c^2 \left(\frac{\partial^2 u_1}{\partial x^2} + \frac{\partial^2 u_2}{\partial x \partial y}\right) - \left(\frac{\partial^2 u_1}{\partial t^2}\right) - \omega^2 u_1 = 0, \tag{15}$$

$$-2\omega x \left(\frac{\partial^2 u_2}{\partial y \partial t}\right) + 2\omega y \left(\frac{\partial^2 u_2}{\partial x \partial t}\right) - \omega^2 y^2 \left(\frac{\partial^2 u_2}{\partial x^2}\right) + \omega^2 y \left(\frac{\partial u_2}{\partial y}\right) + 2\omega^2 xy \frac{\partial^2 u_2}{\partial x \partial y} + \omega^2 x \left(\frac{\partial u_2}{\partial x}\right)$$
$$- \omega^2 x^2 \left(\frac{\partial^2 u_2}{\partial y^2}\right) + c^2 \left(\frac{\partial^2 u_2}{\partial y^2} + \frac{\partial^2 u_1}{\partial x \partial y}\right) - \left(\frac{\partial^2 u_2}{\partial t^2}\right) - \omega^2 u_2 = 0. \tag{16}$$

If we set ω=0, equations 15 and 16 take the form of wave equation in 2-D in which the medium is at rest. For the rest of this paper, we use various forms of these equations for wave propagation.

## III. BREAKING RECIPROCITY WITH CORIOLIS MEAN FLOW

In this section we first overview acoustic reciprocity in a medium at rest, and then discuss numeric results in systems with Coriolis mean flow.

### A. Reciprocity in Acoustics

Consider two single frequency sources in a general space such as the one depicted in Fig. 3. We cut out infinitesimally small volume around two sources. By changing which source is active and inactive successively, we get two acoustic fields, named A and B. By assuming the medium to be at rest, we can as such conclude that the velocity field is irrotational. Thus we can express $\mathbf{u} = (u_1, u_2)$ as the gradient of a scalar, i.e. velocity potential; $\mathbf{u} = \nabla\Phi$. Using Green's second identity for two fields A and B, we have:

$$\int_V (\Phi_A \nabla^2 \Phi_B - \Phi_B \nabla^2 \Phi_A) = \oint_{\partial V} (\Phi_A \nabla \Phi_B - \Phi_B \nabla \Phi_A) \hat{\mathbf{n}}, \tag{17}$$

in which V is the volume of interest and $\partial V$ is the surface boundary of the volume. Velocity potential fields are due to two different sources but the same frequency, so for both potentials we have:

$$\nabla^2 \Phi_A + k\Phi_A = 0, \qquad \nabla^2 \Phi_A + k\Phi_A = 0, \tag{18}$$
$$P_A = -j\omega\Phi_A, \qquad P_B = -j\omega\Phi_B, \tag{19}$$

in which $k = \omega^2/c^2$. Eq. (18) makes the left hand side of eq. (17) vanish. Then by using $\mathbf{u} = \nabla\Phi$ and Eq. (19), we have:

$$\oint_{\partial V} (P_A \mathbf{u}_B \cdot \hat{\mathbf{n}} - P_B \mathbf{u}_A \cdot \hat{\mathbf{n}}) = 0. \tag{20}$$

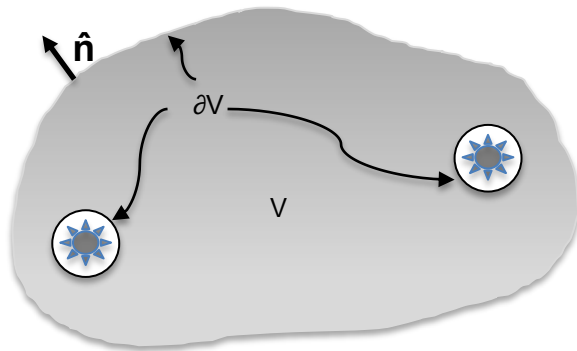

**Figure3: Two sources which are active and inactive successively. We cut out infinitesimally small volume around the two sources and investigate the sound field s.**

This integral vanishes on areas of the surface $\partial V$ where we have perfectly rigid boundary (that is to say $\mathbf{u}\cdot\hat{n}=0$), pressure release boundary (i.e. $P = 0$) or a boundary in which the pressure is impedance times normal velocity ($P = z\mathbf{u}\cdot\hat{n}$). Also if the boundary is far from the sources and we have some absorption in the system, then since the area increases as $r^2$ and the intensity $P\mathbf{u}$ decreases faster than $1/r^2$, the integral approaches zero. In these situations, what is left is the infinitesimal surface areas around the sources. This now becomes reciprocity relation.

## B. Breaking Reciprocity in Acoustics with Coriolis Mean Flow

In deriving reciprocity relation, we assumed that the medium is irrotational and so we could define velocity potential. In the case of having a moving medium with the velocity of $\boldsymbol{V} = r\omega\hat{\theta}$, we cannot define velocity potential, so we are not able to use the aforementioned proof for reciprocity. Nevertheless, it is not guaranteed that the reciprocity is broken. Here, we show by an example that the reciprocity does not generally hold in such a system.

Consider two concentric circles with radii 5 and 40 cm and hard walls as depicted in Fig.4. We select two points on the horizontal axis for our reciprocity analysis distanced 15 and 30 cm from the origin. These two points serve as source and measurement point interchangeably similar to the reciprocity proof we discussed before. Subsequently we numerically solve two different fields A and B corresponding to different source points 1 and 2. Since we use single frequency wave, we can assume $u_1(x,y,t) = U_1(x,y)e^{i\widetilde{\omega}t}$ so eq. (15) and (16) can be rewritten as:

$$-2i\omega\widetilde{\omega}x\left(\frac{\partial U_1}{\partial y}\right) + 2i\omega\widetilde{\omega}y\left(\frac{\partial U_1}{\partial x}\right) - \omega^2 y^2\left(\frac{\partial^2 U_1}{\partial x^2}\right) + \omega^2 y\left(\frac{\partial U_1}{\partial y}\right) + 2\omega^2 xy\frac{\partial^2 U_1}{\partial x \partial y} + \omega^2 x\left(\frac{\partial U_1}{\partial x}\right)$$
$$- \omega^2 x^2\left(\frac{\partial^2 U_1}{\partial y^2}\right) + c^2\left(\frac{\partial^2 U_1}{\partial x^2} + \frac{\partial^2 U_2}{\partial x \partial y}\right) + \widetilde{\omega}^2 U_1 - \omega^2 U_1 = 0, \quad (21)$$

$$-2i\omega\widetilde{\omega}x\left(\frac{\partial U_2}{\partial y}\right) + 2i\omega\widetilde{\omega}y\left(\frac{\partial U_2}{\partial x}\right) - \omega^2 y^2\left(\frac{\partial^2 U_2}{\partial x^2}\right) + \omega^2 y\left(\frac{\partial U_2}{\partial y}\right) + 2\omega^2 xy\frac{\partial^2 U_2}{\partial x \partial y} + \omega^2 x\left(\frac{\partial U_2}{\partial x}\right)$$
$$- \omega^2 x^2\left(\frac{\partial^2 U_2}{\partial y^2}\right) + c^2\left(\frac{\partial^2 U_2}{\partial y^2} + \frac{\partial^2 U_1}{\partial x \partial y}\right) + \widetilde{\omega}^2 U_2 - \omega^2 U_2 = 0. \quad (22)$$

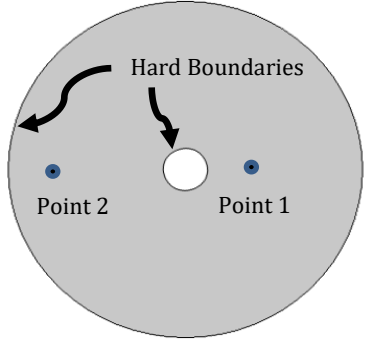

**Figure 4: Acoustic field is analyzed in two concentric circles with radii 5 and 40 cm and hard walls. Points 1 and 2 serve as source point one at a time to produce two different fields A and B. When one point is the source, the other is the measurement point.**

Now we solve eq. (21) and (22) numerically using coefficient form PDE package of COMSOL Multiphysics software. In order to numerically add source points to the system, first we add two small circles of radius 2 mm around points 1 and 2. We then use point source option of the software to add delta functions on the right. The value for this source term is varied until we achieve the same average pressure on the small circles for fields A and B. The process is as follows; first we make the point 1 a point source with some arbitrary value and then we solve numerically eq. (21) and (22). Now that $\boldsymbol{u}$ is found on the whole domain we can find pressure for the entire domain including the small circle. Eq (4) in frequency domain contains $P$ and its derivatives with respect to $x$ and $y$. But since $\boldsymbol{u}$ is already found everywhere, we can use eq. (5) and (6) to replace derivative terms in eq. (4) and calculate $P$ as:

$$P = \frac{i\rho}{\widetilde{\omega}}\left[(i\widetilde{\omega}\omega y - \widetilde{\omega}\omega x - \omega^2 x)U_1 + (-i\widetilde{\omega}\omega x - \widetilde{\omega}\omega y - \omega^2 y)U_2 + (c^2 - \omega^2 y^2)\frac{\partial U_1}{\partial x} + (c^2 - \omega^2 x^2)\frac{\partial U_2}{\partial y} + \omega^2 xy\left(\frac{\partial U_1}{\partial y} + \frac{\partial U_2}{\partial x}\right)\right] \quad (20)$$

This pressure is then averaged on the small circle around the source point. Next, we set point 2 as the source point with some value and we repeat the process and calculate the pressure. Unless we did an educated guess or with the stroke of luck, this pressure is not the same as the previous case. We can linearly adjust the point source value to match these two pressures. At the end of this process, we are guaranteed that $P$ at source points are the same for both fields A and B, so by measuring the velocity $\boldsymbol{u}$ at the other point, we can verify the reciprocity in the system.

In our simulation, for the medium fluid, we used air at room pressure and temperature. The excitation frequency is set to 7 kHz. In the case of no flow where $\omega = 0$ we set the pressure at the source point to be 1 Pa (~94 dB). The velocity numerically calculated at measurement points are the same up to the fourth digit; $U_{point1}/U_{point2} = 1.0000$. This, unsurprisingly, verifies that the reciprocity exist when there is no flow. However, in the case of fluid flow of $\omega = 900$ rad/s with the same pressure at the source point, we have very different velocities; $U_{point1}/U_{point2} = 1.7567$. This shows numerically that the reciprocity is broken in systems in which we have circular flow. Numerical results for both no-flow and with circular flow is depicted in Fig. 5.

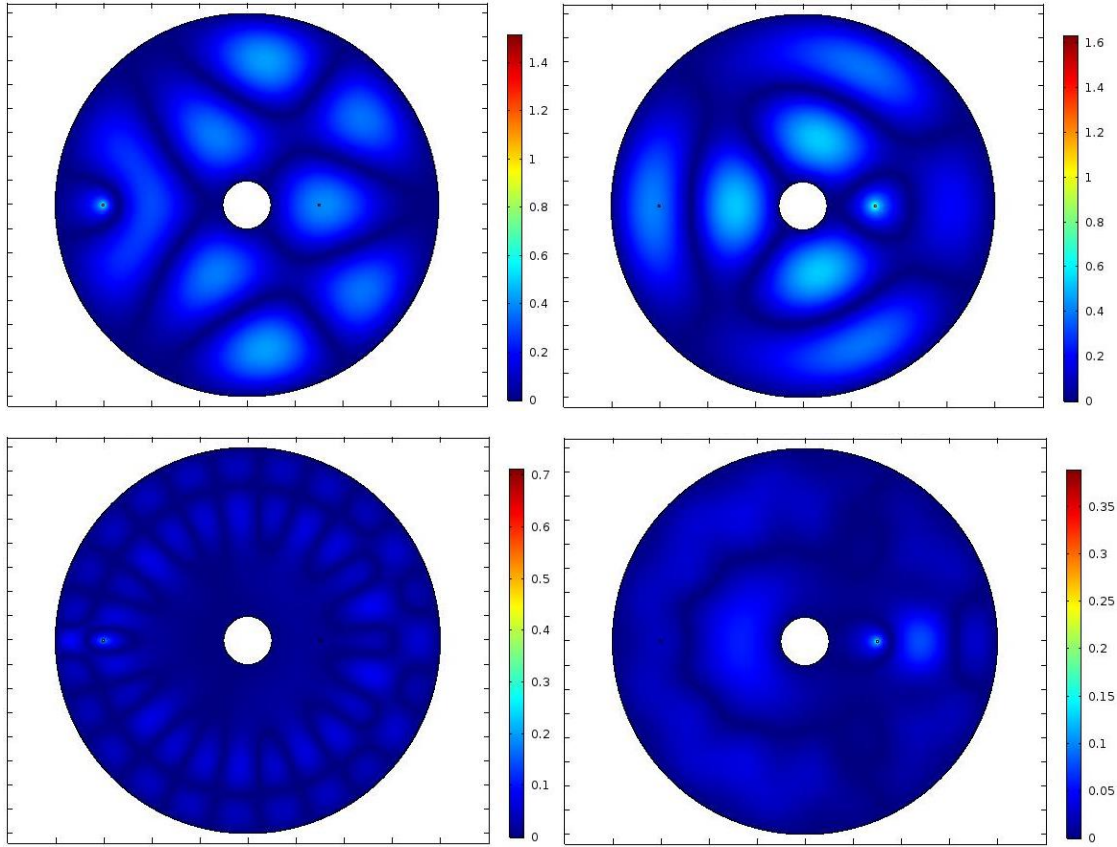

**Figure 5 : Numerical results for the absolute value of velocity U. The no-flow situation is depicted on top and the case of fluid flow of ω = 900 rad/s is on the bottom. Point 1 is the source on the right figures while point 2 is the source point on the left.**

## IV. WAVE PROPAGATION IN THE PRESENCE OF CORIOLIS MEAN FLOW

In this section we numerically investigate wave propagation in a system similar to the one depicted in Fig. 2. In order to see how the wave propagates in time, we numerically solve eq. (15) and (16) using COMSOL Multiphysics software. All the walls are set to be rigid ($u$ =0) except for a quarter (0-90°) of the inner circle where the wall serves as the excitation source. The wave enters from this one quarter for one period of a sine wave. The first and last 1% part of the one period is smoothed out using an exponential function ($e^{-1/x}$) so that all time derivatives exist to avoid numerical artifacts. The medium is air at room pressure and temperature. The simulation is done for one wave length of a sine wave at three excitation frequencies; 4, 7 and 10 kHz. The air rotation speed varies from 0-1000 rad/s in the increments of 100. As it can be seen in Fig. 6, when ω=0 the wave propagates with no deviation from the symmetry axis which is at 45°. While for ω=700 the wave front deviates from the axis of symmetry.

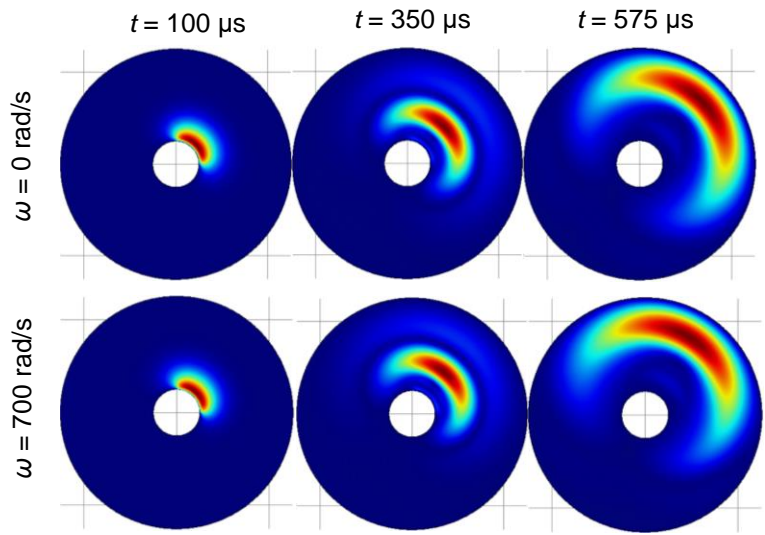

**Figure 6: Numerical results for propagation of a one period sine wave in non-moving (top) and moving (bottom) medium.**

To measure this deviation from the 45° angle, we take a snapshot of the wave pulse at 580 μs. At this moment, the wave front is about 20 cm away from the source and has not reached the outer circle. We locate the center of the area of the top 10% of the wave pulse and measure its angle deviation from the 45°. In order to assess our metric we measure the center of area of the top 30% of the wave pulse as well. The results are depicted in Fig. 7. As evident in this graph, there is not much difference between the two metrics of top 10% and 30%. This graph shows that, as expected, the deviation and non-reciprocity increases with the rotation speed, it also depends on the frequency of the wave. This is to some extend due to the distortion in the wave pulse; longer wave lengths distorts more in this field.

## V. CONCLUDING REMARKS

We presented that to break acoustic reciprocity we can envision a device in which the fluid rotates like a rigid body. In such a system, rotational velocity enters the differential equation of the system as a cross product term with the sound particle velocity. We investigated the effects of frequency of the wave and the rotation speed on the nonreciprocal wave propagation in such system. A device using this phenomena can be used as a filter and/or diode, however, to achieve high level of non-reciprocity, the rotation speed has to be meaningfully high.

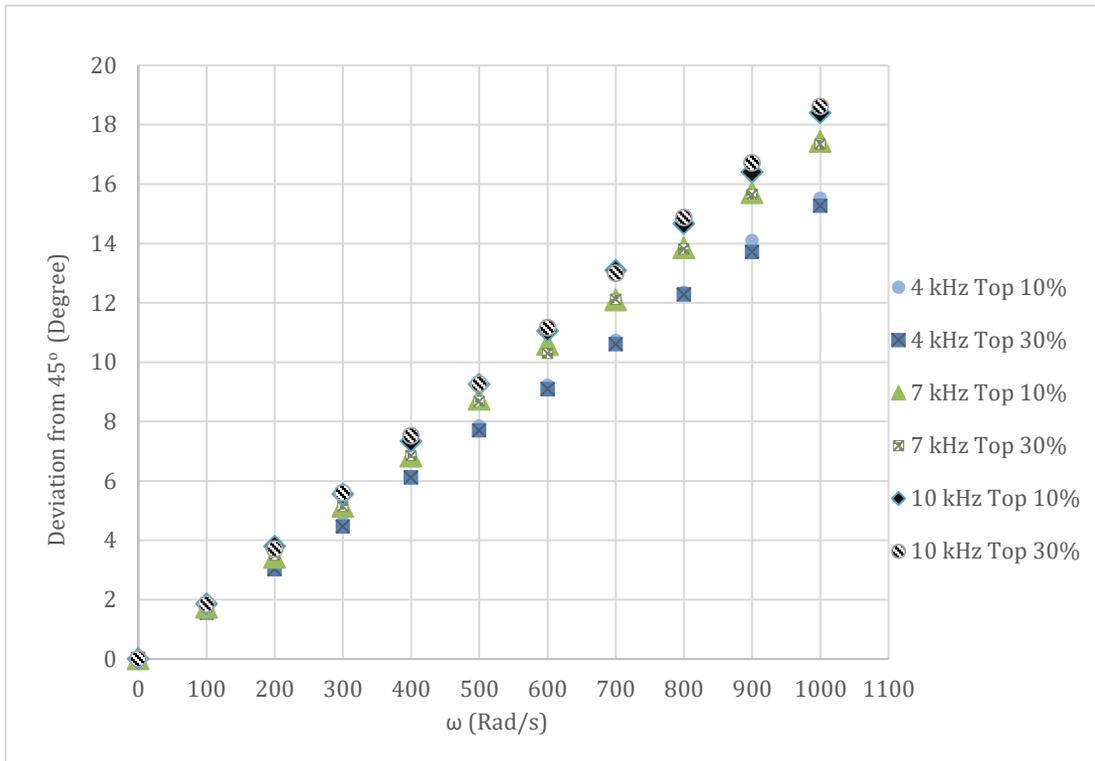

**Figure 7: Numerical results of deviation from the symmetry axis when the wave pulse is 20 cm away from the source and the excitation frequencies are 4, 7 and 10 kHz. Deviation is measured as the angle of the center of the area of the top 10% and 30% of the wave pulse relative to the symmetry axis.**

# REFERENCES

1. Helmholtz, H.v., *Theorie der Luftschwingungen in Röhren mit offenen Enden.* Journal für die reine und angewandte Mathematik, 1860. **57**: p. 1-72.
2. Strutt, J., *Some general theorems relating to vibrations.* Proceedings of the London Mathematical Society, 1871. **1**(1): p. 357-368.
3. Lyamshev, L. *A question in connection with the principle of reciprocity in acoustics.* in *Soviet Physics Doklady.* 1959.
4. Pierce, A.D., *Acoustics: an introduction to its physical principles and applications.* 1991: Acoustical Society of America Melville, NY.
5. Landau, L.D. and E.M. Lifshitz, *Fluid mechanics.* Vol. 6. 1987: Elsevier.
6. Godin, O.A., *Reciprocity and energy theorems for waves in a compressible inhomogeneous moving fluid.* Wave Motion, 1997. **25**(2): p. 143-167.
7. Godin, O.A., *Reciprocity and energy conservation within the parabolic approximation.* Wave Motion, 1999. **29**(2): p. 175-194.
8. Fleury, R., et al., *Sound isolation and giant linear nonreciprocity in a compact acoustic circulator.* Science, 2014. **343**(6170): p. 516-519.
9. Khanikaev, A.B., et al., *Topologically robust sound propagation in an angular-momentum-biased graphene-like resonator lattice.* Nature communications, 2015. **6**.
10. Mansuripur, M., *Classical optics and its applications.* 2002: Cambridge University Press.
11. Kittel, C., *INTERACTION OF SPIN WAVES AND ULTRASONIC WAVES IN FERROMAGNETIC CRYSTALS.* Physical Review, 1958. **110**(4): p. 836-841.
12. Heil, J., B. Luthi, and P. Thalmeier, *NONRECIPROCAL SURFACE-ACOUSTIC-WAVE PROPAGATION IN ALUMINUM.* Physical Review B, 1982. **25**(10): p. 6515-6517.
13. Camley, R.E., *Nonreciprocal surface waves.* Surface Science Reports, 1987. **7**(3-4): p. 103-187.
14. Arntzenius, F., *Time reversal operations, representations of the Lorentz group, and the direction of time.* Studies in History and Philosophy of Modern Physics, 2004. **35B**(1): p. 31-43.
15. Arntzenius, F. and H. Greaves, *Time Reversal in Classical Electromagnetism.* British Journal for the Philosophy of Science, 2009. **60**(3): p. 557-584.
16. Fink, M., *TIME-REVERSAL OF ULTRASONIC FIELDS .1. BASIC PRINCIPLES.* Ieee Transactions on Ultrasonics Ferroelectrics and Frequency Control, 1992. **39**(5): p. 555-566.
17. Fink, M., et al., *Time-reversed acoustics.* Reports on Progress in Physics, 2000. **63**(12): p. 1933-1995.
18. Aharonov, Y. and D. Bohm, *Significance of Electromagnetic Potentials in the Quantum Theory.* Physical Review, 1959. **115**(3): p. 485-491.
19. Berry, M., et al., *Wavefront dislocations in the Aharonov-Bohm effect and its water wave analogue.* European Journal of Physics, 1980. **1**(3): p. 154.
20. Coste, C., F. Lund, and M. Umeki, *Scattering of dislocated wave fronts by vertical vorticity and the Aharonov-Bohm effect. I. Shallow water.* Physical Review E, 1999. **60**(4): p. 4908-4916.
21. Bernal, R., et al., *Normal-Mode\char21{}Vortex Interactions.* Physical Review Letters, 2002. **89**(3): p. 034501.
22. Labbé, R. and J.F. Pinton, *Propagation of Sound through a Turbulent Vortex.* Physical Review Letters, 1998. **81**(7): p. 1413-1416.
23. Howe, M.S., *On the scattering of sound by a vortex ring.* Journal of Sound and Vibration, 1983. **87**(4): p. 567-571.
24. Belyaev, I. and V. Kop'ev, *On the statement of the problem of sound scattering by a cylindrical vortex.* Acoustical Physics, 2008. **54**(5): p. 603-614.
25. Naugolnykh, K., *Sound scattering by a vortex dipole.* The Journal of the Acoustical Society of America, 2013. **133**(4): p. 1882-1884.
26. Nesterenko, V.F., et al., *Anomalous wave reflection at the interface of two strongly nonlinear granular media.* Physical Review Letters, 2005. **95**(15): p. 4.
27. Liang, B., B. Yuan, and J.-c. Cheng, *Acoustic diode: Rectification of acoustic energy flux in one-dimensional systems.* Physical review letters, 2009. **103**(10): p. 104301.

28. Li, B., L. Wang, and G. Casati, *Thermal Diode: Rectification of Heat Flux.* Physical Review Letters, 2004. **93**(18): p. 184301.
29. Boechler, N., G. Theocharis, and C. Daraio, *Bifurcation-based acoustic switching and rectification.* Nature Materials, 2011. **10**(9): p. 665-668.
30. Zhang, J., et al., *Giant nonlinearity via breaking parity-time symmetry: A route to low-threshold phonon diodes.* Physical Review B, 2015. **92**(11): p. 13.
31. Zhu, Y.-F., et al., *Acoustic one-way open tunnel by using metasurface.* Applied Physics Letters, 2015. **107**(11): p. 113501.
32. Maznev, A., A. Every, and O. Wright, *Reciprocity in reflection and transmission: What is a 'phonon diode'?* Wave Motion, 2013. **50**(4): p. 776-784.
33. Farzbod, F. and M.J. Leamy, *Breaking time reversal symmetry with coriolis mean flow systems.* The Journal of the Acoustical Society of America, 2016. **140**(4): p. 3048-3048.